\documentclass[sigconf]{acmart}

\newcommand{\para}[1]{{\vspace{5pt} \bf \noindent #1 \hspace{8pt}}}
\AtBeginDocument{%
  \providecommand\BibTeX{{%
    \normalfont B\kern-0.5em{\scshape i\kern-0.25em b}\kern-0.8em\TeX}}}

\setcopyright{acmcopyright}
\copyrightyear{2020}
\acmYear{2020}
\acmDOI{}

\acmConference[Workshop on Machine Learning in Finance, KDD 2020]{KDD'20 Workshop on Machine Learning in Finance 
}{August 24, 2020}{San Diego, CA, USA}
\acmPrice{}
\acmISBN{}

\begin{document}

\title{Adverse Media Mining for KYC and ESG Compliance}

\author{Rupinder Paul Khandpur}
\email{rupinder.khandpur@moodys.com}
\author{Albert Aristotle Nanda}
\affiliation{%
  \institution{Moody's Analytics}
  \city{New York}
  \state{NY}
}

\author{Mathew Davis}

\author{Chen Li}

\author{Daulet Nurmanbetov}
\affiliation{%
  \institution{Moody's Analytics}
  \city{New York}
  \state{NY}
}
\author{Sankalp Gaur}

\author{Ashit Talukder}
\email{ashit.talukder@moodys.com}
 \affiliation{%
  \institution{Moody's Analytics}
  \city{New York}
  \state{NY}
}

\renewcommand{\shortauthors}{Khandpur and Nanda, et al.}

\begin{abstract}
In recent years, institutions operating in the global market economy face growing risks stemming from non-financial risk factors such as cyber, third-party, and reputational outweighing traditional risks of credit and liquidity.
Adverse media or negative news screening is crucial for the identification of such non-financial risks. 
Typical tools for screening are not real-time, involve manual searches, require labor-intensive monitoring of information sources. 
Moreover, they are costly processes to maintain up-to-date with complex regulatory requirements and the institution's evolving risk appetite.

In this extended abstract, we present an automated system to conduct both real-time and batch search of adverse media for users' queries (person or organization entities) using news and other open-source, unstructured sources of information. 
Our scalable, machine-learning driven approach to high-precision, adverse news filtering is based on four perspectives - relevance to risk domains, search query (entity) relevance, adverse sentiment analysis, and risk encoding. With the help of model evaluations and case studies, we summarize the performance of our deployed application.
\end{abstract}




\settopmatter{printacmref=false}
\maketitle

\section{Introduction}

In today's uncertain geopolitical and social environment, global institutions face growing challenges to their risk management processes arising from Non-Financial Risk (NFR) factors. 
These non-financial risks include, but are not limited to, conduct, cyber, country, compliance, third-party, ESG (Environmental, Social and Corporate Governance) risks. 
Inadequate compliance \& screening controls have cost top banking institutions and other non-financial firms millions of dollars in fines between 2018-2019 alone. In the matter of US Bancorp, fined for lax anti-money laundering controls in 2018~\cite{pete}, it agreed to a \$613 million (USD) settlement with US regulators. It had failed to report suspicious banking activities carried out by the long-time customer, Scott Tucker, from 2011 to 2013, owner of several payday lending businesses. 

With an increased focus on ESG and other regulatory expectations, institutions have realized the importance of integrating adverse news monitoring into their frameworks for managing NFRs.
Adverse Media screening involves the introspection of news and other third-party data sources for potential indicators of negative news associated with an entity (person or company). Adverse media mining makes use of open-source indicators (publicly available information) as essential early warning indicators. 
In a recent study~\cite{barry2019cite,ji2017corporate}, researchers found that Wells Fargo's reputation plummeted after regulators announced the bank's financial fraud. However, Glassdoor reviews signaled the bank had a problem with corporate ethics before the fraud was made public.

The critical challenges for such a screening process include - (1) sheer diversity and volumes of publicly available information, and (2) accurate entity matching and relevance to negative news. Which makes manual monitoring inadequate and may cause lapses in timely access to NFR-related information.
Motivated by this assessment, we outline an automated adverse news screening \& monitoring solution.
The contributions of our work are:
\begin{enumerate}
  \item \textbf{A fast, automated adverse media mining application.} We showcase a system which can scale to high-volume and diverse unstructured data sources, that provides both real-time \& batch processing entity searches for negative news.
  \item \textbf{A high-precision adverse news filtering pipeline.} We develop a novel pipeline to assess the quality of filtered media by its relevance to the risk domains, target entity, and risk attributes (categories and stages of risk).
  \item \textbf{A searchable database of adverse media profiles.} We propose a representation of risk profiles to characterize better, search and retrieve adverse entities.
\end{enumerate}
\begin{figure*}
\includegraphics[width=\linewidth]{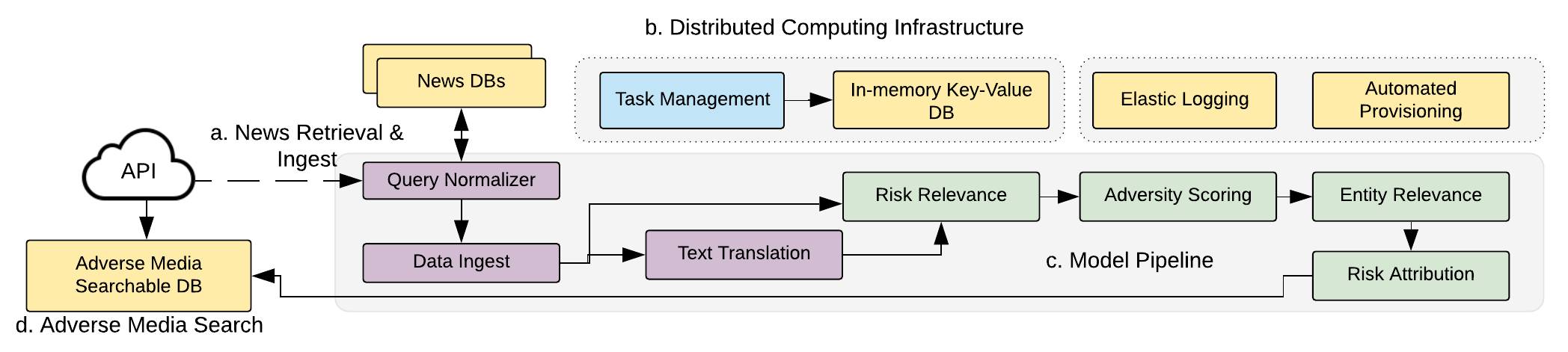}
\vspace{-0.25in}
\caption{A schematic overview of the adverse media mining system.}
\label{fig:systemoverview}
\end{figure*}
\section{System Overview}

We briefly outline the key components of our system as shown Fig.~\ref{fig:systemoverview} for adverse media mining.

\para{a. News Retrieval \& Ingest}
For each new (entity) user query, we query a lucene-powered news database using full-text search to fetch all news articles containing the entity mention. For every single query, we cache the articles and track the search period over which query was issued. This helps the system track and monitor what new news data must be fetched for repeat user queries. All news data is stored in an ElasticSearch database.

\para{b. Distributed Computing Infrastructure}
To process a search query, we make use of Celery-based distributed computing system. We have implemented a ``data-funelling'' pattern of tasks, where each task reduces the number of articles the next task in the pipeline receives. Each task operates on a single article, scheduled using asynchronous work queues. We operate a cluster 18 compute nodes with a master task scheduler that uses an in-memory, key-value data store for book-keeping.

\begin{figure}[h!]
\centering
\includegraphics[height=1.5in]{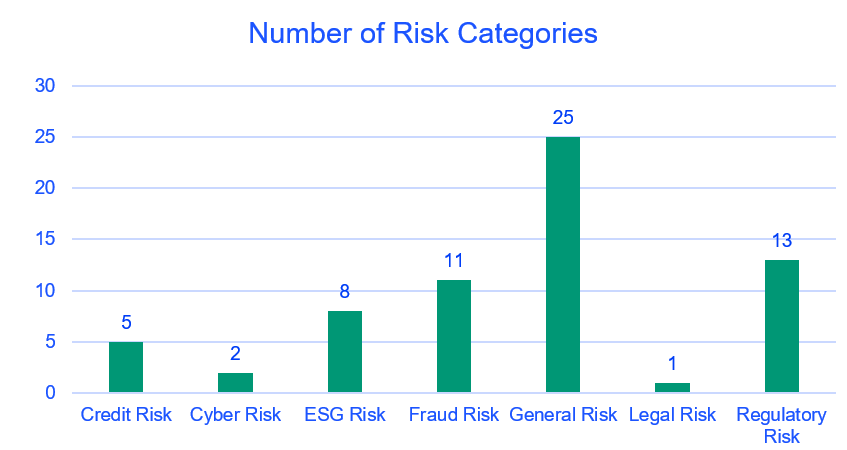}
\caption{Distribution of the 65 risk categories across the 7 risk types.}
\label{fig:riskcategories}
\end{figure}

\para{c. Model Pipeline}
There are five primary models in our pipeline. (1) \textit{Risk Relevance} - a binary relevance classifier based on supervised training of 2500 articles using a support vector machine model. This model classified the relevance of each article to an in-domain (risk-related) or out-domain class. We achieved 0.81 F1 score over a 80/20 train-test split ratio. (2) \textit{Adverse Scoring} - this is a heuristic model for sentiment scoring that relies on the Loughran-McDonald~\cite{loughran2011liability} financial sentiment dictionary for computing adversity score of each article. We further group and weigh differently those subsets of keywords that are negative and related to the legal domain. (3) \textit{Entity Relevance} - Each entity (person or an organization) is assumed relevant to an article if it is an apropos risk domain (compliance). We manually tagged 1200 articles to train a supervised binary logistic regression model using bag-of-words features from contexts extracted around the entity mentions using FlairNLP~\cite{akbik2018coling} using which we achieved F1 score of 0.8 in 80/20 train-test split. (4) \textit{Risk Categorization} -  This step in our pipeline consists of inference risk categories and stage classification. As shown in Fig.~\ref{fig:riskcategories}, we curated a list of fine-grained compliance relevant categories across seven risk types. Using a weakly-supervised labeled data of over 9 Million news documents, we trained CNN-based~\cite{kim2014convolutional} text classifier for the multi-label classification task. The initial set of 65 categories were expanded using sense2vec~\cite{trask2015sense2vec} to query our internal news database further. Our top-3 categorical accuracy was 0.86 in a 70/30 train-test split. (4) \textit{Risk Stage Identification} - We identified five different criminal proceedings stages that typical compliance events might evolve through. Using a similar weak-supervision approach, we trained a multi-class classifier model using XGBoost~\cite{chen2016xgboost} with we achieved a F1 score of 0.93 in a 70/30 train-test split.

\begin{figure}[h!]
\centering
\includegraphics[height=1.6in]{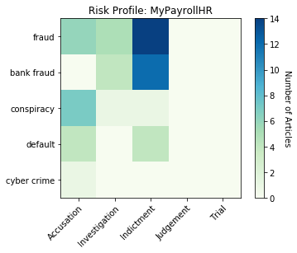}
\caption{A risk profile for an adverse entity ``MyPayrollHR''.}
\label{fig:mypayrollhr}
\end{figure}

\section{Discussion and Conclusion}
Compliance-related risks are continuously evolving. 
Adverse news provides a snapshot essentially in time that can help profile such non-financial risk. For example, in Fig.~\ref{fig:mypayrollhr}, a risk profile for a US-based payroll company MyPayRollHR charged with fraud can be visualized using compliance risk categories and stages with which we can provide actionable insights for the user that best meet their risk appetite. This system is deployed in production in Compliance Catalyst~\cite{comcat}, a KYC monitoring tool.
\bibliographystyle{ACM-Reference-Format}
\bibliography{sample-base}
\end{document}